\documentclass[twoside]{article}
\usepackage{fleqn,espcrc2}
\usepackage{graphicx}

\newcommand{\AmS}{{\protect\the\textfont2
  A\kern-.1667em\lower.5ex\hbox{M}\kern-.125emS}}
\usepackage{here}
\def\beq{\begin{equation}}
\def\eeq{\end{equation}}
\def\bea{\begin{eqnarray}}
\def\eea{\end{eqnarray}}
\def\bq{\begin{quote}}
\def\eq{\end{quote}}

\def\nnb{\nonumber}
\def\ga{\left(}
\def\dr{\right)}

\def\rar{\rightarrow}
\def\lrar{\Longrightarrow}
\def\llar{\Longleftarrow}
\def\nnb{\nonumber}
\def\la{\langle}
\def\ra{\rangle}
\def\nin{\noindent}
\def\ba{\begin{array}}
\def\ea{\end{array}}
\def\bm{\overline{m}}

\def\b{\bullet}

\title{\bf{Light Quark Masses 99}
\thanks{Review talk given at the QCD 99 Euroconference (Montpellier 7-13th July 1999) and plenary talk given at
the QCD Confinement 2000 (Osaka 7-10th March 2000).}}
\author{ Stephan Narison\address{
Laboratoire de Physique Math\'ematique,
Universit\'e de Montpellier II
Place Eug\`ene Bataillon,
34095 - Montpellier Cedex 05, France and KEK Tsukuba, Ibaraki, 350801 - Japan.\\
E-mail:
narison@lpm.univ-montp2.fr}}
\begin{document}
\pagestyle{empty}
\begin{abstract}
\noindent
I give a short historical and a critical review of the determinations of light quark masses from
QCD at dawn of the next millennium. QCD spectral sum rules combined with ChPT give, to order $\alpha_s^3$,
the world average for the running masses: $\overline{m}_s$(2 GeV)= $(118.9\pm 12.2)$ MeV,
$\overline{m}_d(2~\mbox{GeV})=(6.3\pm 0.8)$ MeV, $\overline{m}_u(2~\mbox{GeV})=(3.5\pm 0.4)$ MeV and
the corresponding values of the invariant masses given in Eq. 24. Lower and upper bounds
derived from the positivity of spectral moments are presented in Tables 2 and 3. For a comparison, we critically
review the recent lattice results (section 8 and Table 5) and attempt to deduce the present {\it QCD grand average}
determination (Table 6): 
$\overline{m}_s$(2 GeV)= $(110.9\pm 8.8)$ MeV, to be used with a great care. Then,
we deduce the value: $B_6^{1/2}-0.45(\rm resp.~0.32)B_8^{3/2}\simeq 1.6\pm 0.4~(\rm resp.~1.1\pm 0.3)$
and the lower bound $1.1\pm 0.2$ (resp. $0.7\pm 0.1$), for the combination of the penguin operators, governing the CP-violating parameters
$\epsilon'/\epsilon$ without (resp. with) the inclusion of the final state interaction effects. The result signals a
possible deviation from the leading $1/N$ prediction by about $1\sim 3\sigma$, which should be tested using
accurate non-perturbative calculations.
\end{abstract}
\maketitle
\section{INTRODUCTION}
\nin
One of the most important parameters of the standard model and chiral symmetry is the light
quark masses. Indeed, they are useful for a much better understanding of the realizations of
chiral symmetry breaking
\cite{WEIN,GASSER,CHPT} and for some eventual explanation of the origin of quark masses in unified
models of interactions \cite{FRITZ}. Within some popular parametrizations of the hadronic matrix
elements \cite{BURAS}, the strange quark mass can also largely influence the Standard Model prediction of 
 the
$CP$ violating parameters
$\epsilon'/\epsilon$, which have been mesured recently
\cite{EPS}.
However, contrary to the leptons, where the physical masses
can be identified with the pole of the propagator \footnote{For a first explicit definition of the
perturbative quark pole mass in the $\overline{MS}$-scheme, see \cite{TARRACH} (renormalization-scheme
invariance) and \cite{NARI} (regularization-scheme invariance).}, the quark masses are difficult to define
because of confinement. Instead, they can be treated as coupling constants of the QCD Lagrangian, where the
notion of the running and invariant masses, which are renormalization scheme and scale dependents,  has  been
introduced
\cite{FNR}. In practice, these masses are conveniently defined within the standard
$\overline{MS}$-scheme. In addition to the determination of the ratios of light quark masses (which
are scale independent) from current algebra
\cite{WEIN}, and from chiral perturbation theory (ChPT), its modern version \cite{CHPT}, a lot
of effort reflected in the literature
\cite{PDG} has been put into extracting directly from the data the running quark masses
using the SVZ \cite{SVZ} QCD spectral sum rules (QSSR) \cite{SNB}, LEP experiments \cite{LEP} and lattice data
\cite{LATT}.
In this talk, I shall review the different determinations from these QCD approaches, by
emphasizing the historical developments of the field.
\section{RUNNING AND INVARIANT LIGHT QUARK MASSES IN QCD}
\nin
It is convenient to introduce the dimensionless coupling $x_i(\nu)\equiv m_i(\nu)/\nu$, where
$\nu$ is the renormalization scheme subtraction constant. The running quark mass is a solution
of the differential equation:
\beq
\frac{d\overline{x}_i}{dt}=(1+\gamma(\alpha_s))\overline{x}_i(t)~:\overline{x}_i(t=0)=x_i(\nu)~.
\eeq
In the $\overline{MS}$-scheme, its solution to order $a_s^3$ ($a_s\equiv \alpha_s/\pi$) is:
\bea
\bm_i(\nu)&=&\hat{m}_i\ga -\beta_1 a_s(\nu)\dr^{-\gamma_1/\beta_1}
\Bigg\{1\nnb\\&+&\frac{\beta_2}{\beta_1}\ga \frac{\gamma_1}{\beta_1}-
 \frac{\gamma_2}{\beta_2}\dr a_s(\nu)~
\nnb\\&+&\frac{1}{2}\Bigg{[}\frac{\beta_2^2}{\beta_1^2}\ga \frac{\gamma_1}
{\beta_1}-
 \frac{\gamma_2}{\beta_2}\dr^2-
\frac{\beta_2^2}{\beta_1^2}\ga \frac{\gamma_1}{\beta_1}-
 \frac{\gamma_2}{\beta_2}\dr\nnb\\&+&
\frac{\beta_3}{\beta_1}\ga \frac{\gamma_1}{\beta_1}-
 \frac{\gamma_3}{\beta_3}\dr\Bigg{]} a^2_s(\nu)\nnb\\
&+& 1.95168a_s^3\Bigg\}~,
\eea
where the $a_s^3$ term comes from \cite{VELT}; $\gamma_i$ are the ${\cal{O}}(a_s^i)$ coefficients of the 
quark-mass anomalous dimension, which read
for three flavours:
\beq
\gamma_1=2~,~~~~\gamma_2=91/12~,~~~~\gamma_3=24.8404~.
\eeq
The invariant mass $\hat{m}_i$ has been introduced for the first time by \cite{FNR} in connection
with the analysis of the breaking of the Weinberg sum rules by the quark mass terms in QCD.
\section{RATIOS OF LIGHT QUARK MASSES}
\begin{figure}[hbt]
\includegraphics[width=7cm]{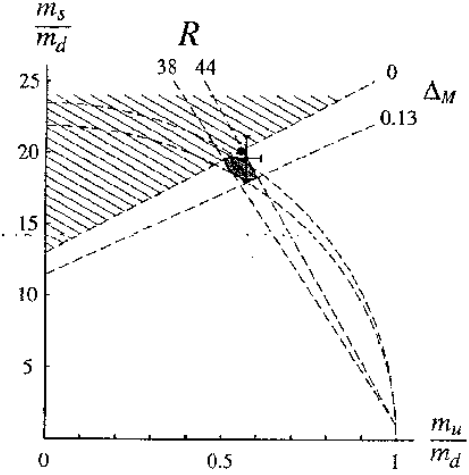}
\caption{$m_s/m_d$ versus $m_u/m_d$ from \cite{LEUTO}. }
\label{fig:largenenough}
\end{figure}
\nin
The ratios of light quark masses are well-determined from current algebra \cite{WEIN}, and ChPT
\cite{CHPT}. In this approach, the meson masses are expressed using a systematic expansion in terms of
the light quark masses:
\bea
 M^2_{\pi^+}&=& (m_u+m_d){ B}  +{\cal O}(m^2)+...\nnb\\
  M^2_{K^+}&=& (m_u+m_s){ B}  +{\cal O}(m^2)+...\nnb\\
  M^2_{K^0}&=& (m_d+m_s){ B}  +{\cal O}(m^2)+...
\eea
where {$ B\equiv -\la\bar\psi\psi\ra/f^2_K$ from the Gell-Mann, Oakes, Renner 
relation \cite{GMOR}}:
\beq
m^2_\pi f^2_\pi\simeq -(m_u+m_d)\la\bar\psi\psi\ra +{\cal O}(m^2)~.
\eeq
 However, only the ratio, which is scale independent can be   well
determined. To leading   order in $m$ \cite{LEUTO}\footnote{In Generalized ChPT, the contribution of the
$m^2$-term can be as large as the $m$ one \cite{GCHPT}, which modifies drastically these ratios.}:
\bea
\frac{m_u}{m_d}& \approx&\frac{M^2_{\pi^+}-M^2_{K^0}+M^2_{K^+}}
{M^2_{\pi^+}+M^2_{K^0}-M^2_{K^+}}\approx  0.66\nnb\\
\frac{m_s}{m_d}&\approx&\frac{-M^2_{\pi^+}+M^2_{K^0}+M^2_{K^+}}
{M^2_{\pi^+}+M^2_{K^0}-M^2_{K^+}}\approx  20
\eea
Including the next order + electromagnetic corrections, the ratios of masses are constrained on the ellipse:
\beq
\ga\frac{m_u}{m_d}\dr^2+\frac{1}{Q^2}\ga\frac{m_s}{m_d}\dr^2=1
 \eeq
where: $Q^2\simeq (m^2_s-\hat{m}^2)/(m^2_d-m^2_u)=22.7\pm 0.8$ using the value of the
$\eta\rar\pi^+\pi^-\pi^0$ from the PDG average \cite{PDG}, though this value can well be in the range
22--26, to be compared with the Dashen's formula \cite{DASHEN} of 24.2; $\hat{m}\equiv (1/2)(m_u+m_d)$. 
In Fig. 1,
one shows the range spanned by $R\equiv (m_s-\hat{m})/(m_d-m_u)$ and the corrections to the GMO 
mass formula $\Delta_M:~M^2_8=(1/3)(4M^2_K-m^2_\pi)(1+\Delta_M)$. The Weinberg mass ratio \cite{WEIN} is
also shown which corresponds to the Dashen's formula  and $R\simeq 43$. At the intersection of
different ranges, one deduces \cite{LEUTO}:
\label{chpt}
\bea
&&\frac{m_u}{m_d}= 0.553\pm 0.043~,~\frac{m_s}{m_d}= 
18.9\pm 0.8,\nnb\\
&&\frac{2m_s}{(m_d+m_u)}= 24.4\pm 1.5.
\eea
The possibility to have a $m_u=0$ advocated in \cite{MANO} appears to be unlikely as it
implies too strong flavour symmetry breaking and is not supported by the QSSR results
from 2-point correlators of the divergences of the axial and vector currents, as will be shown in the
next sections.
\section{QCD SPECTRAL SUM RULES}
\subsection{Description of the method}
\nin
Since its discovery in 79 \cite{SVZ}, QSSR has proved to be a
powerful method for understanding the hadronic properties in terms of the
fundamental QCD parameters such as the QCD coupling $\alpha_s$, the (running)
quark masses and the quark and/or gluon QCD vacuum condensates.
The description of the method has been often discussed in the literature,
where a pedagogical introduction can be, for instance, found in the book \cite{SNB}. In
practice (like also the lattice), one starts the analysis from the two-point correlator:
\beq
\psi_H(q^2) \equiv i \int d^4x ~e^{iqx} \
\la 0\vert {\cal T}
J_H(x)
\ga J_H(0)\dr ^\dagger \vert 0 \ra ~,
\eeq
built from the hadronic local currents $J_H(x)$, which select some specific quantum numbers.
However, unlike the lattice which evaluates the correlator in the Minkowski space-time,
one exploits, in the sum rule approaches, the analyticity property of the
correlator which obeys the well-known K\"allen--Lehmann dispersion relation:
\beq
\psi_H (q^2) = 
\int_{0}^{\infty} \frac{dt}{t-q^2-i\epsilon}
~\frac{1}{\pi}~\mbox{Im}  \psi_H(t) ~ + ...,
\eeq
where ... represent subtraction points, which are
polynomials in the $q^2$-variable. In this way, the $sum~rule$
expresses in a clear way the {\it duality} between the integral involving the 
spectral function Im$ \psi_H(t)$ (which can be measured experimentally), 
and the full correlator $\psi_H(q^2)$. The latter 
can be calculated directly in the
QCD Euclidean space-time using perturbation theory (provided  that
$-q^2+m^2$ ($m$ being the quark mass) is much greater than $\Lambda^2$), and the Wilson
expansion in terms of the increasing dimensions of the quark and/or gluon condensates which
 simulate the non-perturbative effects of QCD. 

\subsection{Beyond the usual SVZ expansion}
\nin
Using the Operator Product Expansion (OPE) \cite{SVZ}, the two-point
correlator reads:
$$
\psi_H(q^2)
\simeq \sum_{D=0,2,...}\frac{1}{\ga q^2 \dr^{D/2}} 
\sum_{dim O=D} C(q^2,\nu)\la {\cal O}(\nu)\ra~,
$$
where $\nu$ is an arbitrary scale that separates the long- and
short-distance dynamics; $C$ are the Wilson coefficients calculable
in perturbative QCD by means of Feynman diagrams techniques; $\la {\cal O}(\nu)\ra$
are the quark and/or gluon condensates of dimension $D$.
In the massless quark limit, one may expect
the absence of the terms of dimension 2 due to gauge invariance. However, it has been
emphasized recently \cite{ZAK} that  the resummation of the large order terms of the
perturbative series, and the effects of the higher dimension condensates due e.g. to instantons, can
be mimiced by the effect of a tachyonic gluon mass $\lambda$ which generates an extra $D=2$ term not present in
the original OPE. Its presence might be understood from the analogy with the short distance linear part
of the QCD potential \footnote{Some evidence of this term is found from the lattice analysis of the static
quark potential \cite{BALI}, though the extraction of the continuum result needs to be clarified.}. The
strength of this short distance mass has been estimated from the
$e^+e^-$ data to be
\cite{SNI,CNZ}:
\beq
\frac{\alpha_s}{\pi}\lambda^2\simeq -(0.06\sim 0.07) ~\rm{ GeV}^2,
\eeq
which leads to the value of the square of the (short distance) string tension:
$
\sigma \simeq -\frac{2}{3}{\alpha_s}\lambda^2\simeq [(400\pm 20)~\rm{ MeV}]^2
$
in an (unexpected) good agreement with the lattice result \cite{TEPER} of about
$[(440\pm 38)~\rm{ MeV}]^2$.
In addition to Eq. 5, the strengths of the vacuum condensates having dimensions $D\leq 6$ are also under
good control, namely:
\begin{itemize}
\item $\la\bar ss\ra/\la\bar dd\ra\simeq 0.7\pm 0.2$ from the meson \cite{SNB} and baryon systems
\cite{JAMI2};
\item $\la\alpha_s G^2\ra \simeq (0.07\pm 0.01)~\rm{GeV}^4$ from
sum rules of $ e^+e^-\rar$ I=1 \rm{hadrons} \cite{SNI} and {heavy quarkonia} \cite{SNH,BB,YNDU},
and from the lattice \cite{DIGI2};
\item $g\la\bar{\psi}\lambda_a/2\sigma^{\mu\nu}G^a_{\mu\nu}\psi\ra\simeq (0.8\pm
0.1)~{\rm GeV}^2\la\bar \psi\psi\ra,$ from
the baryons \cite{HEID,JAMI2} and
the heavy-light mesons \cite{SNhl};
\item $\alpha_s  \la\bar uu\ra^2\simeq
5.8 \times 10^{-4}~\rm{ GeV}^6$ from
$~e^+e^-\rar I=1~ \rm{hadrons}$ \cite{SNI};
\item $g^3\la G^3\ra\approx 1.2~\rm{GeV}^2\la\alpha_s G^2\ra $ from
 { dilute gaz instantons
}~\cite{NSVZ}.
\end{itemize}
\subsection{Spectral function}
\nin
In the absence of the complete data, 
the spectral function is often parametrized
using the ``na\"{\i}ve" duality ansatz:
\bea
&&\frac{1}{\pi}~\mbox{Im}  \psi_H(t)\simeq 2M_H^{2n}f_H^2 \delta (t-M_H^2)+\nnb\\
&& \rm{``QCD
~continuum"}
\times \theta(t-t_c)~, 
\eea
which has been tested \cite{SNB} using $e^+e^-$ and $\tau$-decay data, to give a good description of the
spectral integral in the sum rule analysis; $f_H$ (analogue to $f_\pi$) is
the hadron's coupling to the current ; $2n$ is the dimension of the
correlator; while $t_c$ is the QCD continuum's threshold. 

\subsection{Form of the sum rules and optimization procedure}
\nin
Among the different sum rules discussed in the literature \cite{SNB}, we shall be
concerned with the:\\
$\bullet$ {\bf Laplace sum rule (LSR)} \cite{SVZ,NR,BB}:
\beq\label{usr}
{\cal L}_n(\tau)
= \int_{0}^{\infty} {dt}~t^n~\mbox{exp}(-t\tau)
~\frac{1}{\pi}~\mbox{Im} \psi_H(t)~.
\eeq
The advantage of the Laplace sum 
rules with respect to the previous dispersion relation is the
presence of the exponential weight factor, which enhances the 
contribution of the lowest resonance and low-energy region
accessible experimentally. For the QCD side, this procedure has
eliminated the ambiguity carried by subtraction constants,
arbitrary polynomial
in $q^2$, and has improved the convergence of
the OPE by the presence of the factorial dumping factor for each
condensates of given dimensions. 
As one can notice, there are ``a priori" two free external parameters $(\tau,
t_c)$ in the analysis. The optimized result will be (in principle) insensitive
to their variations. In some cases, the $t_c$-stability is not reached due to the
too na\"{\i}ve parametrization of the spectral function. One can either fix the 
$t_c$-values by the help of FESR (local duality) or improve the
parametrization of the spectral function by introducing threshold effects fixed by
chiral perturbation theory, ..., in order to restore the $t_c$-stability of the
results. The results discussed below satisfy these stability criteria.\\
$\bullet$ {\bf $\tau$-like sum rules} \cite{BNP}--\cite{BRODSKY}:
\beq
{\cal R}_{n}^m= \int_{0}^{M^2_\tau} {dt}~t^n~\ga1-\frac{t}{M_\tau}\dr^m
~\frac{1}{\pi}~\mbox{Im} \psi_H(t)~,
\eeq
The advantage of the $\tau$-like sum rule is the presence of the threshold factor which gives
a zero near the real axis where QCD is not expected to be applicable. Optimal results should be
insensitive to the changes of $M_\tau$, and to the values of the degrees ($m,~n$) of the
moments.\\  
$\bullet$ {\bf Finite Energy Sum Rule (FESR)} \cite{FESR}--\cite{GORI}:
\beq
{\cal M}_n(t_c)
= \int_{0}^{t_c} {dt}~t^n~
~\frac{1}{\pi}~\mbox{Im} \psi_H(t)~,
\eeq
The advantage of the FESR is the separation (to leading order in $\alpha_s$) of the terms of given
dimensions, which gives a set of local duality constraints. However, unlike the two formers, FESR
is sensitive to the high-energy tails of the spectral integral and needs an accurate treatment of
this region, in order that the optimal results are insensitive to the changes of $t_c$.
\section{UP AND DOWN RUNNING MASSES}
\subsection{Pseudoscalar sum rules}
\begin{table}[hbt]
\setlength{\tabcolsep}{0.pc}
\caption{ $(\overline{m}_u+\overline{m}_d)$(1) in MeV 	updated to order $\alpha_s^3$ }
\begin{tabular}{c c c }
\hline 
 && \\
$(\overline{m}_u+\overline{m}_d)$(1) & Sources& Authors\\  
&&\\
\hline
&& \\
  {\bf Best estimate}   
 && \\
$12.8\pm 2.5$&$\pi$ +ChPT&{         BPR95  \cite{BPR,PRADES}}\\
 && \\
   $12.1\pm 2.4$&BPR+&{      CNZ99    \cite{CNZ}}\\
&tachyonic gluon&\\
 && \\
   $12.6\pm 3.2$&$\la\overline{\psi}\psi\ra$& {DN98$^*$    \cite{DOSCH}}\\ 
   &N, $B^*$-$B$ LSR&\\
&$D\rar K^* l\nu$ LSR&\\
 {\bf Others}   
 && \\
$13.2\pm 4.4$&$\pi$+moments&{     Y97    \cite{YND}}\\
 && \\
  $15.6\pm   3.4$& LSR& { SN89 \cite{SN89} 
}\\
&$\pi+\pi'$ NWA&\\ 
   &&\\
\hline
&&\\
      $13.1\pm 1.5$(stat)&$\pm$ 1.3(syst)&{\bf Average}  \\
&&\\
\hline
\end{tabular}
 \\ \thanks{* to order $\alpha_s$ and not included in the average.}
\end{table}
\begin{table}[hbt]
\begin{center}
\setlength{\tabcolsep}{0.4pc}
\caption{Lower bounds on $\overline{m}_{u,d,s}$(2) in MeV to order $\alpha_s$}
\begin{tabular}{c c c }
\hline 
 && \\
Observables& Sources& Authors\\ 
&&\\
\hline
&&\\
{ $\overline{m}_u+\overline{m}_d$}&&\\
$8$&$\pi,~\sigma$  &{           LRT97\cite{LEL}}\\
 7.3&$\pi$&        {   Y97\cite{YND} }\\
         7.&$\la\overline{\psi}\psi\ra$+GMOR& {         DN98\cite{DOSCH}}\\
&&\\
{ $\overline{m}_d-\overline{m}_u$}&&\\
 1.5&$K\pi$&        {  Y97\cite{YND}         }\\
&&\\
{ $\overline{m}_s$}&&\\
         $100$&$K$&        { LRT97\cite{LEL}        }\\
         104&$K$&        {  Y97\cite{YND}        }\\
         90&$\la\overline{\psi}\psi\ra$+ChPT& {         DN98\cite{DOSCH}}\\
&& \\
\hline
\end{tabular}
\end{center}
\end{table}
\begin{table}[hbt]
\begin{center}
\setlength{\tabcolsep}{0.5pc}
\caption{Upper bounds on $\overline{m}_{u,d,s}$(2) in  MeV from \cite{DOSCH} (resp.\cite{SNM99}) to order
$\alpha_s$ (resp. $\alpha_s^3$)}
\begin{tabular}{c c c }
\hline 
 && \\
Observables& Sources& Authors\\ 
&&\\
\hline
&&\\
{ $\overline{m}_u+\overline{m}_d$}&&\\
         11.5&$\la\overline{\psi}\psi\ra$+GMOR& {         DN98\cite{DOSCH}}\\
&&\\
{ $\overline{m}_s$}&&\\
         148&$\la\overline{\psi}\psi\ra$+ChPT& {         DN98\cite{DOSCH}}\\
          $147\pm 21$ &$e^+e^-+\tau$-decay&        {  SN99\cite{SNM99}        }\\

&& \\
\hline
\end{tabular}
\end{center}
\end{table}
\nin
$\bullet$ {\bf Values of $(\overline{m}_u+\overline{m}_{d})$} have been
extracted for the first time in
\cite{BECCHI,NR} using the sum rule of the 2-point correlator associated to the pseudoscalar
current:
\beq
\partial_\mu A^\mu(x)=(m_i+m_j):\bar u(i\gamma_5)d:~.
\eeq
 The analysis has been improved (or disproved) later on by many groups
\cite{GASSER,SNB}, \cite{ER}--\cite{YND}, by the inclusion of higher order terms or/and by a more
involved parametrization of the spectral function (threshold effects, ChPT,...).
However, this channel is quite peculiar due to the Goldstone nature of the pion, where the value
of the sum rule scale ($1/\tau$ for Laplace and
$t_c$ for FESR) is relatively large of about 2 GeV$^2$ compared with the pion mass, where the
duality between QCD and the pion is lost. This implies an important role of the higher states
(radial excitations or/and theoretical parametrizations of the spectral function above the 3$\pi$
threshold) in the analysis, and then led to some controversial results, which hopefully can be
cured by the presence of the new $1/q^2$
\cite{ZAK,CNZ} due to the tachyonic gluon mass, which enlarges the duality region to lower scale
and then minimizes the role of the higher states into the sum rule. The errors due to the QCD part
of the sum rules, which is now known to order $\alpha_s^3$, are much less than from the parametrization of
the spectral function. 
Among the
available results, we consider that the best estimates of
$(\overline{m}_u+\overline{m}_{d})$ from  this channel come from \cite{BPR} ($\pi$+ ChPT
parametrization of the $3\pi$ continuum) and from \cite{CNZ} (inclusion of the tachyonic gluon mass into the 
analysis of \cite{BPR}). The result of \cite{BPR} to order $\alpha_s^2$ has been extended to order
$\alpha_s^3$ by \cite{PRADES}. Also the result of \cite{BPR} 
updates the one in \cite{DR87}. Ref.
\cite{YND} uses the positivity of the higher state contributions plus the  moment inequalities to order
$\alpha_s$. In
\cite{SN89}, one treats the
$\pi'$ in a Narrow Width Approximation (NWA), while the QCD expression is to order $\alpha_s^2$. These different
determinations are quoted in Table 1, after including into these published results the perturbative contributions
of the known 
$\alpha_s^3$-term \cite{CPS97}. The effect of this term is quite small as the PT series converges quite well at
the sum rule working region of about 1.5 GeV. Indeed, the PT expression of e.g. the Laplace sum rule
normalized to ${(\overline{m}_u+\overline{m}_{d})^2}$ reads:
\beq
{\cal L}\sim 1+4.82a_s+21.98a_s^2+53.14a_s^3+{\cal O}(a_s^4).
\eeq
One can notice the good consistency of the results from the different forms of the sum rules in the
pion channel.
We give in
Table 1 the average of these updated determinations,
where we have added an extra 10\% error
which takes into account the systematics of the approach\footnote{The estimation of the systematic error is based
on the (un)ability of the method for reproducing the hadron masses and couplings
\cite{SVZ,SNB}.}. 
\\
$\bullet$ {\bf Lower bounds for
$(\overline{m}_u+\overline{m}_{d})$} based on moments inequalities and the
positivity of the spectral functions have been obtained, for the first time,
in \cite{BECCHI,NR}. These bounds have been rederived recently in
\cite{LEL,YND} to order $\alpha_s$\footnote{The inclusion of the $\alpha_s^3$ term will decrease by about 10\%
the strength of these bounds, which is within the expected accuracy of the result.}.
Their optimal values quoted in Table 2 
exclude the low value of
about 6 MeV  given by \cite{GUPTA}. 
\subsection{Scalar sum rules}
\nin
$\b$ {\bf Lower bounds on $(\overline{m}_{d}-\overline{m}_{u})$} have been extracted for the
first time in
\cite{PAVER} using the sum rule of the 2-point correlator associated to the scalar
current:
\beq
\partial_\mu V^\mu(x)=(m_d-m_u):\bar d(i)u:~,
\eeq
which is sensitive to leading order to the quark-mass difference. The analysis has been extended later
on by many authors \cite{SNB,DR87,YND}. However, the analysis relies heavily on the less controlled
nature of the
$a_0(980)$, where its $\bar qq$ nature appears to be favoured by the present data \cite{MONT}. In
the $I=0$ channel, the situation of the $\pi$-$\pi$ continuum is much more involved due to the
possible gluonium nature of the low mass and wide $\sigma$ meson \cite{VENEZIA,HAD99,MONT}. Instead,
one can also use these sum rules the other way around, i.e. by using the values of the quark masses
from the pseudoscalar sum rules and their ratio from ChPT, in order to test the nature of these
resonances
\cite{SNB,HAD99}. 
\subsection{Direct extraction of $\la\bar\psi\psi\ra$}
\nin
$\b$ {\bf The chiral condensate} can be directly extracted from the sum rules (nucleon, $B^*$-$B$
splitting, vector form factor of $D^*\rar K^* l\nu$), which are particularly sensitive to it and
to the mixed condensate
$\la\bar\psi\sigma^{\mu\nu}(\lambda_a/2)G^a_{\mu\nu}\psi\ra\equiv M^2_0\la\bar\psi\psi\ra$. A global
fit from these different channels gives, to order $\alpha_s$, the running condensate value at 1 GeV \cite{DOSCH}:
\beq
0.6\leq \la\bar\psi\psi\ra/[-225~\rm{MeV}]^3\leq 1.5,
\eeq
a result also recovered by the lattice \cite{VALDIKAS}.\\
$\b$ {\bf Lower and
upper bounds} on the light quark masses given in Tables 2 and 3, can be obtained by transforming this
result using the PCAC relation in Eq. 5,
and the positivity of the ${\cal O}(m^2)$ term. These results are independent on how
chiral symmetry is realized (ChPT or generalized ChPT ?).
\subsection{$\overline{m}_{u,d,s}$ to order $\alpha_s^3$ from sum rules + ChPT}
\nin
One should note from Table 1 the consistency of the results from the pion channel and the one
from the direct extraction of $\la\bar\psi\psi\ra$, which is an a posteriori support of the
validity of the OPE for the $\pi$-sum rule in the working region, and signals the absence of the large
effects  due to instantons, which may break the OPE. 
Using the ratios from ChPT in section (\ref{chpt}),
one can deduce in units of MeV, the value of the running masses at 2 GeV to order $\alpha_s^3$ given in
Table 6. We have used the conversion factor:
\beq
\overline{m}_i(1)\simeq  (1.38\pm 0.06)\overline{m}_i(2),
\eeq
 for running, to order $\alpha_s^3$, the
results from 1 to 2 GeV, which corresponds to the average value of the QCD scale
$\Lambda_3\simeq (375\pm 50)$ MeV from PDG \cite{PDG} and others \cite{BETHKE}. I remind that the
errors in these determinations already take into account the systematics of the method (see
Table 1).
\section{DIRECT EXTRACTIONS OF $\overline{m}_s$ }
\begin{table}[hbt]
\begin{center}
\setlength{\tabcolsep}{0.pc}
\caption{Direct extractions of $\overline{m}_s$(1) in MeV to order $\alpha_s^3$.
}
\begin{tabular}{c c c c}
\hline
              &&& \\
 Channels&$\overline{m}_s$ (1)&Sources& Authors\\ 
&&&\\
\hline
&&&\\
Kaon SR
              &$165\pm 15$& &               {                 SN89$^{*}$ \cite{SN89}}\\
              &$155\pm 25$& &               {                 DPS99 \cite{DPS99}}\\
           &$\bf 155\pm 25$& &\bf Largest Range\\
              &&&\\
Scalar SR
              &$203\pm 20$& &               {                 CPS97 \cite{CPS97}}\\
              &$143\pm 17$& &               {                 CFNP97 \cite{CFNP97}}\\
              &$160\pm 30$& &               {                 J98 \cite{J98}}\\
              &$159\pm 11$& &               {                 M99 \cite{M99}}\\
           &$\bf 175\pm 48$& &\bf Largest Range\\
              &&&\\
$            (              \tau$-like $\phi$ SR
              &$173\pm 33$&$R_\phi$ &               {                 SN95,99
\cite{SNM95,SNM99}}\\
              $e^+$-$e^-$ data               &$176\pm 31$&$\Delta_{10}$ &              
{                 }\\
              $+\tau$-decay            )              &$186\pm 31$&$\Delta_{1\phi}$ &              
{                }\\
           &$\bf 179\pm 39$& &\bf Largest Range\\

              &&&\\
$\Delta S=-1$ 
              &$234^{+61}_{-76}$& &               {                 Aleph99 \cite{LEP}}\\
               part of&$200\pm 50$&&               {                 CKP98 \cite{CKP98}}\\
              $\tau$-decay&$164\pm 33$& &               {                 PP99 \cite{PP99}}\\
           &$\bf 213\pm 82$& &\bf Largest Range              \\
&&&\\ 
\hline
&&&\\
\bf Average of & \bf ~Largest &\bf ~Ranges &\boldmath $166.7\pm 18.8$\\
&&&\\
\hline
\end{tabular}
\end{center}
\end{table}
\subsection{Pseudoscalar sum rules}
\nin
In the strange quark channel, we quote in Table 4 the results from \cite{SN89} and
\cite{DPS99}, and consider the largest range spanned by these previously quoted results. We
consider that this conservative range already takes into account and may even overestimate the
systematics of the method. One should notice here that, unlike the case of the pion, 
the result is less
sensitive to the contribution of the higher states continuum due to the relatively 
higher value of $M_K$, though the parametrization of the spectral function still gives larger errors
than the QCD series.
\subsection{Scalar sum rules}
\nin
Following the pioneer's analysis of \cite{PAVER}, $m_s$ has been obtained by different
authors \cite{CDPS95}--\cite{M99} by using the
$K\pi$ phase shift data for parametrizing the spectral function. The different values 
obtained from this channel to order $\alpha_s^3$ is given in Table 4. Like in the case of the
pseudoscalar channel, the errors are dominated by the uncertainties for parametrizing the spectral
function.
\subsection{$m_s$ from $e^+e^-+\tau$-decay data}
\nin
One can combine the $e^+e^-\rar I=0,~1$ hadrons and the rotated recent $\Delta S=0$ 
component of the $\tau$-decay data in
order to extract $m_s$. Unlike previous sum rules, one has the advantage to have a
complete measurement of the spectral function in the region covered by the analysis. 
We shall work with:
 \bea
R_{\tau,\phi}&\equiv&\frac{3|V_{ud}|^2}{2\pi\alpha^2}S_{EW}\int_0^{M^2_\tau}
ds\ga 1-\frac{s}{M^2_\tau}\dr^2\nnb\\ &&\ga 1+\frac{2s}{M^2_\tau}\dr\frac{s}{M^2_\tau}
\sigma_{e^+e^-\rar \phi,\phi',...}~,
\eea
and the $SU(3)$-breaking combinations \cite{SNM95,SNM99}:
\beq
\Delta_{1\phi}\equiv R_{\tau,1}-R_{\tau,\phi},
~~~~\Delta_{10}\equiv R_{\tau,1}-3R_{\tau,0}~,
\eeq
which vanish in the $SU(3)$ symmetry limit;
$\Delta_{10}$ involves the difference of the isoscalar ($R_{\tau,0}$) and isovector ($R_{\tau,1}$)
sum rules \`a la Das-mathur-Okubo \cite{DMO}. The PT series converges quite well at
the optimization scale of about 1.6 GeV \cite{SNM99}. E.g, normalized to $\overline{m}^2_s$, one has:
\beq
\Delta_{1\phi}\sim 1+\frac{13}{3}a_s+30.4a_s^2   +(173.4\pm 109.2)a_s^3.
\eeq
It has been argued in
\cite{M98} that
$\Delta_{10}$ can be affected by large $SU(2)$ breakings, but this claim has not been confirmed from the result
based on the other sum rules not affected by these terms \cite{SNM99}. The largest range of values
from different form of the sum rules is given in Table 4, which one can compare with the average
of $(178\pm 33)$ MeV given in \cite{SNM99}. An
upper bound deduced from the positivity of $R_{\tau,\phi}$ is given in Table 3.
\subsection{$m_s$ from the $\Delta S=-1$ part of $\tau$-decay}
\nin
One can also extract $m_s$ from the Cabibbo suppressed channel of $\tau$-decay
\cite{LEP,CKP98,PP98,PP99}, using different $\tau$-like moments. Unlike the case of the neutral
$\phi$-meson current, where the QCD series is more convergent, here the convergence is quite bad,
such that one needs to select an appropriate combination (spin 1+0 pieces) for obtaining an
acceptable result. Though a complete agreement has been obtained in the previous analysis of
\cite{PP98} with the two other determinations \cite{LEP,CKP98}, a recent analysis in \cite{PP99} is
lower and more precise than the former, though still in agreement with the previous ones due to the
generous errors given there. Ref.
\cite{PP99} argues that one should consider the previous results as an upper bound rather than a
determination, an argument which needs to be confirmed. By inspecting the results in \cite{PP99}, we
notice that the estimate decreases with increasing power of moments, rather than stabilizing.
Therefore, it can be more appropriate to consider the conservative range spanned by the results from
the three moments which is
$(180\pm 68)$ MeV, rather than taking their average quoted in Table 4 from \cite{PP99}. This
conservative value is in better agreement with the two other determinations. It is also interesting
to notice that the results from $\tau$-decay are in good agreement with the one from
$e^+e^-$ data, an agreement which is a priori expected because of the similarity of the two
approaches. 
\subsection{$m_s$ and $m_s/(m_u+m_d)$ from QSSR}
\nin
More generally, one can notice from Table 4 that there is a total agreement of the sum rule
results from different channels. As already
mentioned, we expect that the largest ranges given in Table 4 already include the QSSR systematics. Using the
average of these largest ranges, one can deduce the {\it pure} QSSR determination of $m_s$, 
to order $\alpha_s^3$ given in Table 4 and evolved to 2 GeV in Table 6.
Combining this result with the sum rules determination of $(m_u+m_d)$ in Table 1 (section (5)),
one can deduce {\it pure QSSR} prediction of $m_s/(m_u+m_d)$ in Table 6, which is
in agreement with and an independent test of the ChPT result $24.4\pm 1.5$ given in section (3), though less
accurate. 
\section{QSSR + ChPT FINAL RESULTS}
\nin
We take the average of $m_s$ from $m_u+m_d$ (Table 1) + the ChPT ratio and from
the direct determination in Table 4. Then, we obtain the {\it final average} from QSSR+ChPT to order
$\alpha_s^3$ in Table 6.
Combined with the ratios from ChPT in section 2, this value leads to the  values of $m_{u,d}$ given in
Table 6. As already discussed in previous sections, the quoted error already include the
systematics of the methods. The size of the error is
within the expected accuracy of the sum rule results. Using Eq. 2, it is
trivial to extract the value of the invariant mass
$\hat{m}_i$. One obtains in units of MeV:
\label{invariant}
\bea
\hat{m}_u&=&3.9\pm 0.7~,~~~~{\hat{m}_d= 7.1\pm 0.8 }\nnb\\
\hat{m}_s&=&133.3\pm 18.8~,
\eea
where the error is larger than the corresponding running mass due to $\Lambda$ in the evolution procedure.
\section{COMPARISON WITH THE LATTICE}
\subsection{Lattice approaches for/by non-experts}
\nin
One usually starts from the QCD action and partition function:
\beq
 Z=\int {\cal D} A_\mu~ det~{ M }
~{ e^{\int d^4x \ga -\frac{1}{4} G^{\mu\nu}G_{\mu\nu}\dr}}
\eeq
 integrated over gauge field configurations. The fermion contributions are included
into the non-local $det~ M$ term. For the analysis, one works like in the sum rule
approach, with the 2-point correlator defined in previous sections, which is
saturated by the intermediate states $|n \ra\la n|$. In this way, the two-point
correlator can be expressed as:
\bea
&&\sum_x\la 0\vert 
J(x)
\ga J(0)\dr ^\dagger \vert 0 \ra 
=\nnb\\
&& \sum_n  
\la 0\vert 
J(x)\vert n\ra\la n\vert
\ga  J(0)\dr ^\dagger \vert 0 \ra \frac{e^{-E_n{ t}}}{ 2E_n}
\eea
where the zero momentum states $E_n$ tend to the masses $M_n$ of the resonances. In
the (ideal) asymptotic limit $t\rar\infty$, the exponential factor kills the effect of the
different excitations, such that the lowest ground state contribution
dominates. In practice, this approximation is expected to be realized
when the splitting between the ground state and the radial excitation is
large enough.
\subsection{Practical limits of the lattice}
Besides the usual statistical and finite size (about 1\% if the lattice size $L\geq 3$
fermi and $m_\pi L\geq$ 6), errors
inherent to the lattice, which can be minimized using modern technology, there are still
large uncertainties related to the uses of field theory on the lattice due to the finite
values of the lattice spacing $a$:\\
$\bullet$ The different operators mix at finite $a$.\\
$\bullet$ The discretization errors specific to each actions, which are ${\cal O}(a)$
for the Wilson (explicit breaking of chiral symmetry ($\chi S$)) and Domain wall
(extra 5th dimension in order to preserve $\chi S$) actions, 
${\cal O}(a^2)$ for the staggered (reduction of quark couplings with high-momenta
gluons) and 
${\cal O}(a\alpha_s)$ for the Clover (inclusion  of the mixed quark-gluon operator)
actions. For typical values of $1/a\approx$ 2 GeV, the error is $\approx$ 10-30\%,
which can be reduced by computing at different values of $a$.\\
$\bullet$ The well-known quenched approximation (no inclusion of the fermion
contribution $ln~Det~ M$), which implies a modification of $\chi S$ with unphysical
singularities for $m_q=0$ or practically for $m_q\leq m_s/3$ (recall that in this
approximation: $M_{\eta'}\approx m_\pi=0~ (\equiv$ large $N_c$-limit)), which induces an
error of about 20\% that can be estimated from  the deviation of the predictions
from the observed meson masses and couplings or/and from the choices of the mesons for
setting the scale (string tension). \\ 
$\bullet$ The extrapolation of the results to light quark masses with the help of the
meson mass dependence expected from ChPT, which for typical values $1/a=2$ GeV, and
keeping $m_\pi L\geq$ 6, one requires $L/a\geq 90$ in order to avoid finite volume
effects. At present, $L/a\approx 32$ (quenched) and $L/a\approx 24$ (unquenched) far below this limit.\\
$\bullet$ The errors due to the matching of the lattice and the continuum at a typical
lattice conversion scale of 2 GeV can be minimized using the non-perturbative
renormalization.
\begin{table}[hbt]
\setlength{\tabcolsep}{0.0pc}
\caption{Lattice Quenched Results up to NNLO since 98.}
\begin{tabular}{c c c c}
\hline 
 &&& \\
Group&$\overline{m}_s$ (2) &Sources& Comments\\ 
&&&\\
\hline
{\bf 98}&&&\\
OHIO\cite{SEATTLE98}&129$\pm$ 23&&{  Staggered }\\
APE \cite{APE98}& 130$\pm 18$&K*&{                  NLO, AWI}\\
&&&{                  Wilson, Clover}\\
&121$\pm 13$&$K,~\phi$&{                  NLO}\\
&&&NPR+AWI\\
{\bf 99}&&&\\
                     CP-PACS\cite{CP-PACS}&143$\pm 6$&$\phi$&{                 
AWI+VWI}\\ &&& Wilson\\
&115$\pm 2$&$K$&{                  qChPT } \\
                     JLQCD\cite{JLQCD}&129$\pm 12$&$\phi$&{                 
AWI+VWI}\\ &&& Kogut-Suss.\\
&106$\pm 7$&$K$&{                  qChPT } \\

                  $\alpha$-UKQCD\cite{A-UKQCD}&97$\pm 4$&$f_K$&{                 
NNLO}\\ 
&&$N$&$\nearrow$ {                  10\%}\\
DESY\cite{DESY}& 105$\pm 4$&K*&{                  NNLO}\\
&&&AWI\\
&&&{               I. Wilson, }\\

BNL\cite{BNL}&95$\pm 26$&$f_\pi, K$& {                  Domain }\\
BNL\cite{BNL2}&$130\pm 21$&&Walls\\
APE\cite{APE99} &114$\pm 9$&Q-Prop.&{NNLO}\\
                    &&&\\
\hline
&&&\\
                 Average&112.9&$\pm 1.5$ (stat.)& $\pm 22.3$ (syst.)\\
&&&\\
\hline
\end{tabular}
\end{table}
\subsection{Lattice results and estimated errors}
\begin{table*}[hbt]
\begin{center}
\setlength{\tabcolsep}{0.2pc}
\newlength{\digitwidth} \settowidth{\digitwidth}{\rm 0}
\caption{ Summary for $\overline{m}_{u,d,s}$(2) in  MeV}
\begin{tabular*}{\textwidth}{@{}l@{\extracolsep{\fill}}rrrrrr}
\hline 
 &&&&&&\\
\bf Sources&$\overline{m}_s$&$2{m_s}/{(m_u+m_d)}$&$\overline{m}_u+\overline{m}_d$
&$m_u/m_d$&$\overline{m}_u$   &$\overline{m}_d$  \\ 
&&&&&&\\
\hline
 &&&&&& \\
\bf QSSR&&&&&&\\
Table 1&$115.8\pm 19.7$&$\llar$ ChPT+&\boldmath $9.5\pm 1.4$&+ChPT $\lrar$&$3.4\pm 0.6$&$6.1\pm 0.8$\\
Table 4&\boldmath$120.8\pm 14.6$&$\lrar$ \boldmath $25.5\pm 4.8$&&&&\\
\bf Average
&\boldmath$118.9\pm 12.2$&+ChPT $\lrar$&$9.8\pm 1.2$&+ChPT $\lrar$&$3.5\pm 0.4$&$6.3\pm 0.8$\\
&&&&&&\\
{\bf LATTICE}&&&&&&\\
Table 5: quenched&~$112.9\pm 22.4~^*$&&less reliable&&&\\ 
$n_f=2$ dynamical \cite{CP2}&$97\pm 11\sqrt{2}~^*$
&&less reliable&&&\\
&&&&&&\\
\bf GRAND AVERAGE
&\boldmath$110.9\pm 8.8$&+ChPT $\lrar$&$9.1\pm 0.9$&+ChPT $\lrar$&$3.2\pm 0.3$&$5.9\pm 0.5$\\
&&&&&&\\
{\bf QSSR BOUNDS}&\boldmath$90\leq...\leq 168$&&\boldmath$7\leq...\leq
12$&&&\\
Tables 2 and 3&&&&&&\\
\hline
\end{tabular*}
\end{center}
\thanks{* Our conservative error estimate}
\end{table*} 
\nin
From the previous discussions, we consider that:\\
$\bullet$ The conservative quenched lattice errors are about 20\%.\\
$\bullet$ The extraction of $m_{u,d}$ is less reliable than $m_s$.\\
Therefore, we shall only consider the value of $m_s$ obtained from the lattice which we
shall compare with the one obtained in previous sections. Lattice results prior 98 have been already
reviewed in \cite{LATT}. The different results for 98 and 99 are given in Table 5 for different
actions, where one can see a large spread of predictions, which with the given errors are
inconsistent each others. We mainly attribute the source of this discrepancy to the underestimate of
the systematic errors given there. Most of these results have been obtained using the
non-perturbative renormalization \cite{MARTI}, and Ward identities for the axial (AWI) and/or vector
(VWI) currents, and constraints from ChPT in the extrapolation procedure. There is also a systematic 
discrepancy between the results from the kaon and $\phi$, $K^*$ or the nucleon  masses, which might
be some indications of the quenching errors, while e.g. the splitting of the $K$-$K^*$ is no longer
resolved.  As one can see from Table 5, the quenched lattice predictions are in the range:
\beq
\overline{m}_s(2)\simeq (69-181)~\mbox{MeV}~,
\eeq
where part of this range is already excluded by the bounds given in Tables 2 and 3.
Instead, one can also quote (to be taken carefully) the na\"{\i}ve average given 
in Table 5 at NNLO \footnote{However,
as discussed in \cite{APE99}, the inclusion of the higher order $\alpha_s$ corrections
obtained in \cite{CHET} in the conversion of the lattice to the continuum results tends to
decrease slightly the value of $m_s$ by about 3\%.}, where we have added our guessed 20\% estimate of the
lattice systematic errors based on the previous comments. At this approximation, where a comparison with the
previous results from QCD spectral sum rules is meaningful, one can notice a surprisingly good agreement. \\
Some attempts to put dynamical fermions have been done in \cite{SESAM}, and more recently with 2 flavours in
\cite{CP2}. In \cite{CP2}, some of the problems encountered in the quenched approximation
(discrepancy between the $\phi$ and $K$ results,...) seems to be resolved. Though promising, the approach is
not enough mature for the different systematics to be fully under control. We quote in Table 6
this result adopting a more conservative error than the original one. 
\section{SUMMARY}
\nin
We have reviewed the different determinations of light quark masses from ChPT,
QCD spectral sum rules (QSSR), $e^+e^-$ and $\tau$-decay data, and compared the one of 
the strange quark mass
with the recent lattice results: \\
$\b$ The sum of light quark masses $m_u+m_d$ to order $\alpha_s^3$ from different QSSR analysis is given in Table
1 and the resulting average value.\\
$\b$ Lower (resp. upper) bounds based on the positivity and analyticity properties of the
spectral functions are given in Table
2 (resp. Table 3).
$\b$ Different direct determinations of the strange quark mass to order $\alpha_s^3$ are compiled in Table 4.  
\\ $\b$ Combined results from these four
methods lead to the final average given in Table 6 to order $\alpha_s^3$, where the errors are typically
the 10\% systematics  of the QSSR approach. An eventual failure of this result should signal a new phenomena not
accounted for in the OPE discussed in this paper.
\\
$\b$ We have compared this final result with the recent (after 98)
lattice determinations (Table 5)
which belong in the range given by Eq. 27 and which lead to the 
average in Table 5. \\
$\b$ Within the present uncertainties of various approaches, we consider that
there is a good agreement between the previous sum rule and lattice results. Attempting
to give the final QCD value, we average the different results
in Table 6, and deduce the {\it QCD Grand
Average} given in this table, to be used carefully. \\
$\b$ However, we expect that a future high-precision measurement of the light quark
masses will be difficult to reach due to the systematic errors inherent to each methods, which, often,
different authors do not include into their results ! 
\\ 
$\b$ Finally, one should remind
that, in the phenomenological analysis, one should use the value of
$m_s$ into the expression of any hadronic matrix elements or/and observables which are known at the same level of
approximation.  This consistency condition is not often respected in the 
literature.
\section{APPLICATION TO $\epsilon'/\epsilon$}

\nin
One of the {\it most fashionable} applications of the previous result is the one to the CP violating
parameters $\epsilon'/\epsilon$,  where $\epsilon'$ is related to $ A[K_L\rar(\pi\pi)_{I=2}]/A[K_S\rar(\pi\pi)_{I=0}]$ and characterizes the (direct)
CP-violation in the decay amplitude of $K\rar\pi\pi$; $\epsilon=A[K_L\rar(\pi\pi)_{I=0}]/
A[K_S\rar(\pi\pi)_{I=0}]$ is
the (indirect) CP-violation from
$K^0$-$\bar{K}^0$ mixing.
It is known \cite{BURAS} that the dominant effects in the analysis of $K_{L,S}\rar (\pi\pi)_{I=0,2}$ amplitudes are
due to the QCD and electroweak penguin operators:
\bea
Q_6&\equiv& \ga \bar s_\alpha d_\beta\dr_{V-A}\sum_{u,d,s}\ga \bar 
\psi_\alpha\psi_\beta\dr_{V+A}\nnb\\
&\approx& B^{1/2}_6/m_s^2+{\cal O}(1/N)\nnb\\
Q_8&\equiv& \frac{3}{2}\ga \bar s_\alpha d_\beta\dr_{V-A}\sum_{u,d,s}\ga e_\psi\bar 
\psi_\alpha\psi_\beta\dr_{V+A}\nnb\\
&\approx& B^{3/2}_8/m_s^2+{\cal O}(1/N)~.
\eea
For $M_t\simeq 165$ GeV and $\Lambda_4\simeq 340$ MeV, 
the {\it simplified} SM prediction without (resp. with) the inclusion of final-state interaction effects
 \cite{PALLANTE}, is\cite{BURAS}:
\bea
 \frac{\epsilon'}{\epsilon}&\approx& 9.75({\rm resp.}~15.34){\rm Im}\lambda_t
\Bigg{[}\frac{110~{\rm MeV}}{\overline{m}_s(2)}
\Bigg{]}^2\nnb\\&\times&\Bigg{[}B^{1/2}_6-0.54(\rm resp. 0.32)B^{3/2}_8
\Bigg{]}.
\eea
where $\lambda_t= V_{td}V_{ts}^*$ is expressed in terms of the CKM matrix elements. 
Using Im $\lambda_t\simeq (1.34\pm 0.30)\times 10^{-4}$ \cite{BURAS,ALI},
and the measured
value \cite{EPS}
$(\epsilon'/\epsilon)_{exp}\simeq (21.4\pm 4.0)\times 10^{-4}$, one can
deduce, from the average value and the lower bound of $m_s$ in Table 6
\footnote{This is an improved estimate of the one in \cite{SNM99}.}:
\bea
B_6^{1/2}-0.54(0.32)B_8^{3/2}&\simeq& 1.6\pm 0.4(1.1\pm 0.3)~,\nnb\\
&\geq& 1.1\pm 0.2( 0.7\pm 0.1), \nnb\\
\eea
which signals a violation of about $1\sim 3\sigma$ for the leading
$1/N$ vacuum saturation prediction \footnote{It has been noticed from various QSSR analysis \cite{SNB} that
the four-quark vacuum condensate also deviates from the vacuum saturation by about a factor 2 to 3.}
: $B_6^{1/2}\approx B_8^{3/2}\approx 1$. This result and the final-state interaction effects should
be tested using more accurate non-perturbative calculations. It is only after performing these tests
that one can make a sharper conclusion on the SM prediction of the $CP$-violation
\footnote{A direct analysis 
can avoid the $m_s$-dependence \cite{GUISTI}.}. 

\section*{QUESTIONS AND DISCUSSIONS}
\nin
Lively general discussions, many questions and some comments
have followed this talk, 
As they have been also addressed to the previous talks in this session, they
have not been reported here. 
\end{document}